\documentclass[%
dvipsnames,
reprint,
 aps,
]{revtex4-1}

\usepackage{graphicx}
\usepackage{dcolumn}
\usepackage{bm}
\usepackage{epsfig}
\usepackage{multirow}
\usepackage{wrapfig}
\usepackage{braket}
\usepackage{xcolor}
\usepackage{appendix}
\usepackage{amsmath}

\begin{document}

\title{Scaling behavior of the momentum distribution of a quantum Coulomb system in a confining potential}

\author{J. A. E. Bonart$^{a,b}$,
W. H. Appelt$^{a}$,  D. Vollhardt$^a$,
L. Chioncel$^{a,c}$
}
\affiliation{$^{a}$ Theoretical Physics III, Center for Electronic
Correlations and Magnetism, Institute of Physics, University of
Augsburg, 86135 Augsburg, Germany}
\affiliation{$^{b}$ BMW AG, 80807 Munich, Germany}
\affiliation{$^c$ Augsburg Center for Innovative Technologies, University of Augsburg,
86135 Augsburg, Germany}

\date{\today}

\begin{abstract}
We calculate the single-particle momentum
distribution of a quantum many-particle system
in the presence of the Coulomb interaction and
a confining potential. The region
of intermediate momenta, where the confining potential dominates, marks a crossover from a Gaussian distribution valid at low momenta to a power-law behavior valid at high momenta.
We show that for all momenta the momentum distribution can
be parametrized by a $q$-Gaussian distribution whose 
parameters are specified by the confining potential.
Furthermore, we find that the functional form of the probability of transitions between the confined ground state 
and the $n^{th}$ excited state is invariant under scaling of
the ratio $Q^2/\nu_n$, where $Q$ is the transferred momentum
and $\nu_n$ is the corresponding excitation energy. 
Using the scaling variable $Q^2/\nu_n$ the maxima of the transition probabilities 
can also be expressed in terms of a $q$-Gaussian. 
\end{abstract}

\maketitle

\section{Introduction}

The single-particle momentum distribution plays an important role in our understanding
of the ground-state properties of quantum many-particle systems~\cite{si.so.89,west.75}. It
is defined as the average number of particles with momentum {\bf k} in an N-particle system,
$n({\bf k})=\langle \Psi|\sum_\sigma a^\dagger_{{\bf k}\sigma} a^{}_{{\bf k}\sigma} |\Psi \rangle$.
Here  the normalized $N$-particle state of the system is represented by $|\Psi \rangle$
and $a^\dagger_{{\bf k}\sigma} (a^{}_{{\bf k}\sigma})$ are the creation
(annihilation) operators for particles with momentum {\bf k} and spin projection
$\sigma$.
In real space the one-particle density matrix
\begin{equation}
    \rho({\bf x}_1, {\bf x}^\prime_1) = \int  \prod_{i=2}^N d{\bf x}_i
\psi^\dagger ({\bf x}_1,{\bf x}_2, ..., {\bf x}_N) \psi({\bf x}^\prime_1,{\bf x}_2, ..., {\bf x}_N)
\end{equation}
measures the change of the $N$-particle wave function when a particle
is moved from ${\bf x}^\prime_1$ to ${\bf x}_1$ while all other particles are fixed.
In homogeneous
systems this two-point function depends only upon the separation:
$ \rho({\bf x}_1, {\bf x}^\prime_1) =\rho(|{\bf x}_1-{\bf x}^\prime_1|)$. Accordingly, the momentum
distribution and the one-particle density matrix are related by Fourier transformation:
\begin{eqnarray}
    n({\bf k}) = \int d{\bf x}_1 \int d{\bf x}^\prime_1
    e^{i{\bf k}\cdot ({\bf x}_1-{\bf x}^\prime_1)} \rho({\bf x}_1-{\bf x}^\prime_1). \label{nofp}
\end{eqnarray}

The momentum distribution, eq.~(\ref{nofp}), is determined by a product of two field operators whose short-distance behavior can be calculated exactly using   renormalization group methods~\cite{wi.zi.72,al.ci.12,va.ry.12,an.bo.10}. 
The latter techniques when applied in nuclear physics 
decouple the low- from the high-momentum degrees of freedom 
and leave the scattering cross section invariant~\cite{west.75,re.si.85,bjor.69,bjor.67,poli.73,poli.74}.
Furthermore, the nuclear momentum distributions
calculated within the impulse approximation~\cite{ch.wi.52,as.wi.52} provide
universal scaling  laws for the high-momentum tails~\cite{west.75} of one- and
two-particle momentum distributions~\cite{al.ci.12,va.co.11}. These tails are the consequence of short
range-correlations in the nuclear wave functions~\cite{pi.wi.92}.
Renormalization group arguments~\cite{bo.ro.12,ho.ba.13} 
have also shown that
high-momentum
tails of momentum distributions factorize into a product between a universal function of momentum, which
is determined by two-particle physics and a factor depending on
the low-momentum
structure of the many-body state~\cite{an.bo.10}. This observation goes back to Kimball~\cite{kimb.73,kimb.75} who pointed out that when two particles are
sufficiently close their interaction dominates, 
and the 
two-particle Schr\"odinger equation provides a reasonable 
starting point to compute quantum mechanical observables
from the knowledge of the pair-wave function.

Experimental measurements of $n({\bf k})$ involve inelastic scattering processes with energy
and momentum transfers larger than the characteristic length scale of the scatterer. They determine the double differential scattering cross section $d^2\sigma/ d\Omega d\omega$ for a given infinitesimal solid angle $d\Omega$ and energy $d\omega$
of the scattered particle, respectively. The incident energy and the scattering angle are fixed during the experiment, and the scattering cross section is measured as a function of the transferred momentum and energy.
The data analysis of the measured scattering cross section generally employs the impulse approximation~\cite{ch.wi.52,as.wi.52}, which assumes that a single particle is struck by the scattering probe, and that the particle recoils freely 
from the collision.
Within the impulse approximation the scattering cross section is proportional to the Compton profile $d^2\sigma/d\Omega d\omega \propto J(k_z)$. The latter
can be calculated directly by integrating the momentum distribution $n({\bf k})$ in a plane perpendicular to the scattering vector $k_z$: $J(k_z) = \int \int n({\bf k}) dk_x \ dk_y$.
The proportionality implies that, whenever  
the measured scattering cross section
is  modelled within the impulse
approximation~\cite{si.so.89,west.75} and is found to be invariant under 
some scaling transformation, the Compton profile will show the same scaling behavior.

In this paper we investigate whether the momentum distribution of a Coulomb system, which yields the Compton profile by integration, 
also shows scaling behavior. In high energy physics it is well known that the scaling of the scattering cross section is a consequence of confinement (``Bjorken scaling'').
Indeed, by assuming the existence of a simple confining potential
for two point particles, Elitzur and Susskind ~\cite{el.su.72} derived the scaling behavior of the
resonance excitations found experimentally in deep inelastic reactions~\cite{bl.gi.70,bjor.69}.
Making use of Kimball's observation~\cite{kimb.73,kimb.75} we will therefore compute the momentum distribution of two interacting electrons by numerically
solving the two-particle Schr\"odinger equation for a repulsive Coulomb
interaction in the presence of a confining potential.
In the following we work in atomic units, where the unit
length is $a_0=1 \ \textrm{Bohr} (0.529167 \times 10^{-10} \ \textrm{m})$,
the unit of mass is the electron mass $m$, and the unit of energy is 1 Hartree
(1 $\textrm{Ha} = %e^2/4\pi \epsilon_0 a_0 =
27.2113$ eV).
To keep the investigation general we consider an algebraic confining potential  of the model form
$V = \alpha \left|x/a_0\right|^\eta$.
%, where $a_0$ is the Bohr radius.
The condition $\eta >0$ ensures that the potential produces bound states, and we chose $\alpha=1\, \textrm{Ha}$.
We will show that the momentum distribution of a quantum many-particle system
interacting via the Coulomb interaction in the presence of
a confining potential can be parametrized by a $q$-Gaussian distribution whose parameters are determined by the confining potential.

In Sec.~\ref{sec:kimbal} we compute the high-momentum tails of the momentum
distribution, eq.~(\ref{nofp}), in the groundstate and show that they obey scaling relations.
A crossover in the momentum density from an ordinary Gaussian
distribution at small momenta to a power-law behavior at larger momenta
occurs when the Coulomb potential dominates the confinement. In the
cross-over region of size $\approx 5\,/a_0$ %
we find that
the shape of the momentum distribution is described
by a $q$-Gaussian with $k$-dependent parameters.
At large momenta we recover the exact results obtained by renormalization group methods~\cite{al.ci.12,va.ry.12,an.bo.10,bo.ro.12,ho.ba.13}.
Using the solutions
of the two-particle Schr\"odinger equation of
Sec.~\ref{sec:kimbal} we show in Sec.~\ref{sec:susskind} that
when the confinement dominates the Coulomb
interaction, the transition matrix elements
into excited states ($n^{th}$-bound level) due to momentum absorption also obey
$q$-Gaussian distributions, and we connect the $q$-parameter
to the shape of the confining potential. The
$q$-Gaussian used to parametrize the momentum distribution is characterized by
parameters which are different from those used to fit the transition probabilities
between the bound states.
Finally, in Sec.~\ref{sec:Discussion} we relate these results to the recent observation~\cite{se.ap.18} that the Compton profiles of all alkali elements can be collapsed onto a single curve which is described by a $q$-Gaussian.

\section{Ground state: Kimball's approach to the momentum distribution}
\label{sec:kimbal}
We consider non-relativistic electrons with Coulomb interaction whose Hamiltonian reads~\cite{kimb.73,kimb.75}:
\begin{equation}\label{ham}
H=-\frac{\hbar^2}{2m}\sum_{i=1}^{N}\nabla_i^2 - \sum_{i=1}^{N} \sum_{I=1}^{N_I} \frac{e^2Z_I}{|{\bf x}_i-{\bf R}_I|}
+ \sum_{i < j} \frac{e^2}{|{\bf x}_i-{\bf x}_j|} \ .
\end{equation}
Here ${\bf x}_i$ and ${\bf R}_I$ are the electronic and nuclear coordinates, respectively, $eZ_I$ is the charge of the
$I^{th}$ nucleus, and $m$ is the mass of an electron.
The first term is the kinetic energy of the electrons,
and the remaining two terms represent the Coulomb repulsion between
the electrons and  nuclei, and between the electrons themselves.
The eigenstates of the Hamiltonian (\ref{ham}) are time independent wave functions
$\psi({\bf x}_1,{\bf x}_2, ..., {\bf x}_N)$ which are normalized to the volume of the system.
Here periodic boundary conditions are assumed and spin indices are suppressed.
Following the idea of Kimball \cite{kimb.73,kimb.75},
the behavior of the wavefunction at large momenta is determined by the dependence at small distances between two particles. As the distance
approaches zero, the dynamics of adjacent particles is dominated by the Coulomb force.

Instead of the explicit electron-nucleon attraction 
we consider an effective confining potential  $V(r)$ whose nature or origin we do not
further specify, since we merely wish to understand the consequences of the confining potentials, e.g., possible
scaling properties.
Introducing relative coordinates ${\bf x}={\bf x}_1-{\bf x}_2$ and center-of-mass coordinates ${\bf X}=({\bf x}_1+{\bf x}_2)/2$ for the two particles as well as the reduced mass $\mu=1/2$, the Schr\"odinger equation becomes
\begin{equation}
    \left( -%\frac{\hbar^2}{2\mu}
    \frac{\partial^2 }{\partial x^2}  +
    V(x)+ %\frac{e^2}{x_r}
    \frac{1}{x}
    \right)\psi_n(x) = %(E_n-H^\prime)
    E_n \psi_n(x) .
    \label{eq:model1}
\end{equation}

\begin{figure}[h]
   \includegraphics[width =0.45\textwidth]{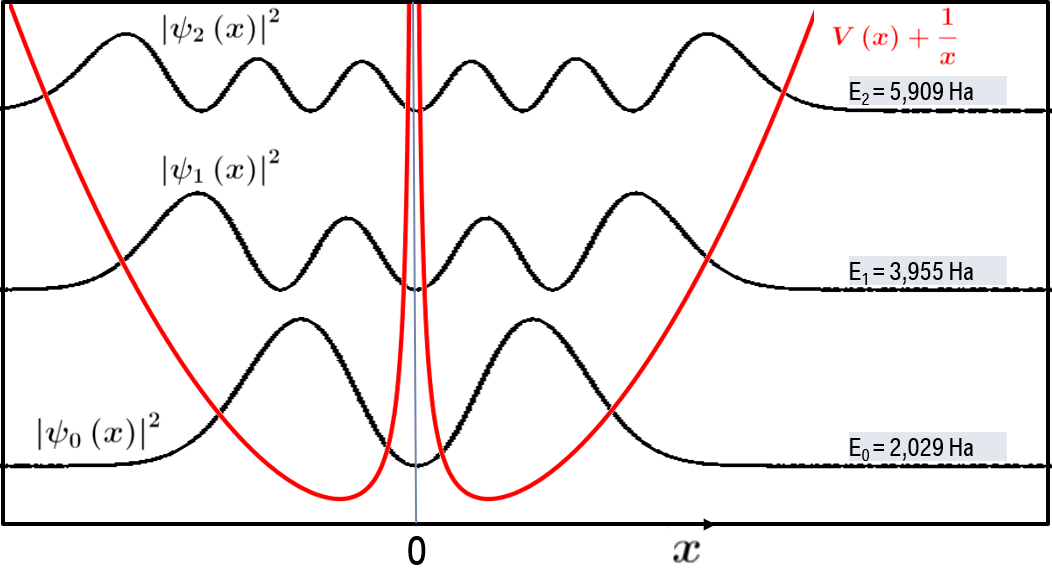}
    \caption{Probability density $|\psi_n(x)|^2$ of the pair wave functions for $n=0,1,2$ obtained from the solutions of eq.~\ref{eq:model1}. Red solid line: Confining potential $V(x)$ together with the repulsive Coulomb term $1/x$, which is singular at $x = 0$. The corresponding energies 
    $E_n$ of the ground state and the first excited states are 
    shown on the right. For a better visibility the probability densities are separated along the vertical axis.} 
    \label{Fig:psi_n}
\end{figure}

The solutions to
eq.~(\ref{eq:model1}) are denoted by $\psi_n(x)$ with the corresponding eigenenergy $E_n$, are shown in 
Fig.~\ref{Fig:psi_n}. 
The ansatz for the total wave function introduced by Kimball~\cite{kimb.73,kimb.75} separates
the dependence on the relative coordinates from that of the
center of mass motion
\begin{eqnarray}\label{separation}
 \Psi_n &\equiv& \Psi_n ({\bf x}_1,{\bf x}_2,{\bf x}_3, ..., {\bf x}_N )  \nonumber \\
    & = & \Psi_n ({\bf x},{\bf X},{\bf x}_3, ..., {\bf x}_N )  \nonumber \\
   &\simeq& \Phi({\bf X},{\bf x}_3, ..., {\bf x}_N )\, \psi_n({\bf x}).
\end{eqnarray}

The one-particle density matrix for the relative coordinates is then given by
\begin{eqnarray}
    \rho({\bf x}, {\bf x}^\prime) & \simeq& \int  d{\bf X} \prod_{i=3}^N d{\bf x}_i
\   \Phi^\dagger ({\bf X},{\bf x}_3, ..., {\bf x}_N) \psi^\dagger_n ({\bf x})  \nonumber  \\
  &  & \Phi^{} ({\bf X},{\bf x}_3, ..., {\bf x}_N) \psi^{}_n({\bf x}^\prime) \nonumber \\
  & = & \overline{\rho}  \ \psi^\dagger_n ({\bf x})  \psi^{}_n ({\bf x}^\prime),
\end{eqnarray}
where $\overline{\rho}$ represents the integral of the
$N$-particle wave function
over all coordinates ${\bf x}_i$ and
${\bf X}$.
From $\rho({\bf x}, {\bf x}^\prime)$ we obtain 
the two-particle correlation function
$g({\bf x})$, which is defined as
$ g({\bf x}) = \rho({\bf x},{\bf x})  =  \overline{\rho}  \ \psi^\dagger_n ({\bf x})  \psi^{}_n ({\bf x})$
and is proportional to the probability
of two particles being at a distance ${\bf x}$.
The momentum distribution is computed according to eq.~\ref{nofp}.

\begin{figure}[h]
\includegraphics[width =0.45\textwidth ]{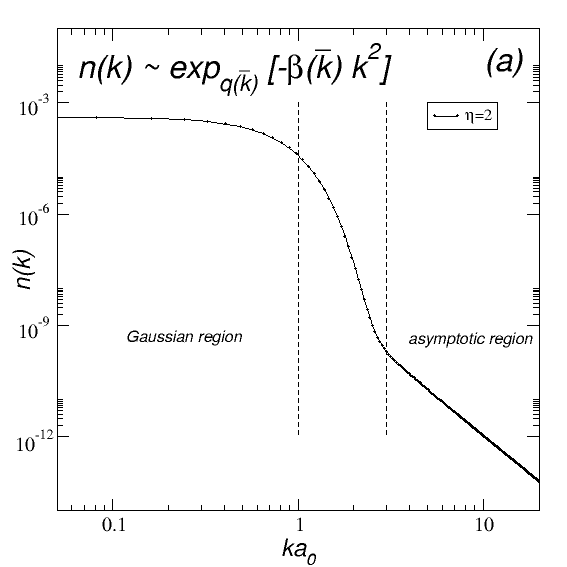} \\ 
\includegraphics[width =0.45\textwidth ]{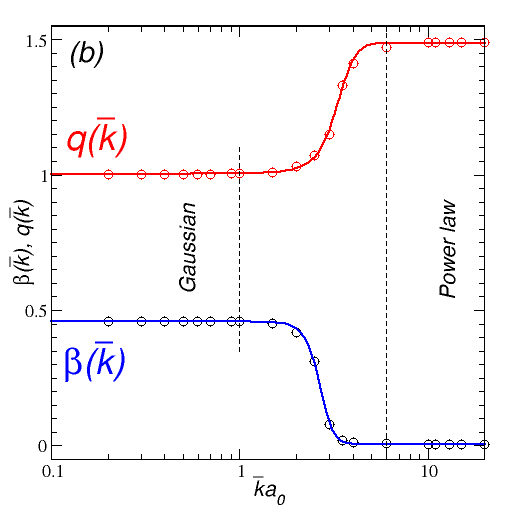}
    \caption{(a) Computed momentum distribution $n(k)$ of electrons in the presence of a confining potential
    ($\alpha=1 \textrm{Ha}$, $\eta = 2$). The momentum distribution is fitted by a $q$-Gaussian in the entire momentum region. 
    (b) Dependence of the $q, \beta$-parameters on the average
    momentum $\bar{k}$.
    The low momentum region shows an ordinary Gaussian behavior ($q=1$). In the intermediate region
    ($k_{min}\approx 1.0\,/a_0 < k< k_{max}\approx5.0\,/a_0$) $q$-Gaussian type of fits are possible with momentum dependent parameters.
    At large momenta the power-law dependence is recovered.
    \label{nk-k-plot}}
\end{figure}

Fig.~\ref{nk-k-plot}~(a) shows the momentum dependence of $n(k)$ for the confining
potential $V(x)=\alpha \left|x \right/a_0|^\eta$ with $\eta =2$, where $k$ is the magnitude of $\bf k$.
One clearly sees a Gaussian regime at small $k$, 
which is followed by a crossover into the asymptotic region at large $k$.
In order to specify the momentum dependence by a unique
functional form we fit the momentum distribution  
with a $q$-Gaussian~\cite{tsal.88}.
To determine the $(q,\beta)$ parameters we collect k values 
into ``bins" that are characterized by 
an average value $\bar{k}$.
The latter quantity is computed from the interval in which the
fitting to the $q$-Gaussian form is performed and which contains at least five
points in the low momentum and an order of magnitude more (fifty) points
in the asymptotic region.  The initial fitting parameters of the $i^{th}$ bin 
are the final parameters of the $(i-1)^{th}$ $\bar{k}$-bin.
For a given bin the same values 
$q(\bar{k}),\beta(\bar{k})$ parametrise 
the momentum dependence as:
\footnote{In this context we note that the ground-state wave function of a quantum particle in a Coulomb potential has the form of a q-Gaussian in momentum space \cite{vi.pl.12}. Furthermore the distribution of the energies of all elements of the periodic table were also observed to follow a $q$-Gaussian~\cite{am.za.10}.
}

\begin{equation}
n_{q(\bar{k}),\beta(\bar{k})}(k) = \frac{1}{\sqrt{2}\beta(\bar{k}) C_{q(\bar{k})}} \exp_{q(\bar{k})}\left(-\beta(\bar{k}) k^2\right). \label{ftsal}
\end{equation}
For arbitrary values of $q$, the $q$-exponential is defined as
$\exp_q(x) = \left[1+\left(1-q\right)x\right]^{1/(1-q)}$.
In eq.~\ref{ftsal} $C_{q(\bar{k})}$ is a normalization constant and $\beta(\bar{k})$ controls
the width of the distribution 

In Fig.~\ref{nk-k-plot}~(b) we plot the momentum dependence of the ($q,\beta$)-parameters.
We see that for low momenta $q(\bar{k})=1$ while $\beta(\bar{k}) = 0.49$. In fact, the
$\exp_q$-function
becomes the exponential function in the limit of $q \rightarrow 1$, whereby the
Gaussian distribution is recovered. The low momentum region  corresponds to large distances. In this case
the Coulomb interaction is negligible and the solutions become plane waves~\cite{kimb.73,kimb.75},
which form Gaussian wave packages leading to a Gaussian momentum dependence.
In the crossover region, i.e., in the range $\bar{k}_{min}\approx 1.0\,/a_0$ to $\bar{k}_{max}\approx 5.0\,/a_0$,
a smooth transition between Gaussian and power law behavior is observed in the
momentum dependence of the ($q,\beta$) parameters.

For large values of $k>k_{max}$ the $q$-Gaussian distribution has a power law dependence $f_{q,\beta}(x)|_{x\rightarrow \infty} \propto x^{2/(1-q)}$, which in our case
amounts to constant values of the parameters $q(\bar{k}) = 1.50$ and $\beta(\bar{k}) = 0.06$.
Thus, at large momenta the power-law behavior is
recovered~\cite{kimb.73,kimb.75}, and the asymptotic behavior of the
momentum density agrees with the result
obtained by the renormalization group
approach~\cite{bo.ro.12,ho.ba.13}.
We note that the $q$-Gaussian fits and the corresponding $q(\bar{k}),\beta(\bar{k})$-parameters differ for different values of the confining potentials (values of the parameter $\eta$). Nonetheless, constant values of $q(\bar{k}),\beta(\bar{k})$ are obtained for the low momentum region as well as the asymptotic region.

In the following section we will show that the transition
matrix elements between the ground state and
the $n^{th}$-energy level caused by absorption
of a momentum, calculated from the solutions of eq.~\ref{eq:model1}, also obey scaling. This result was previously
found by Elitzur and Susskind within a simplified parton model~\cite{el.su.72}.
Here we compute the full transition probability and demonstrate that the scaling
functions for the maxima of the transition probabilities
can also be expressed by $q$-Gaussians.
Since transition probabilities are connected to
the scattering cross section from which the
Compton profiles follow, our calculation proves that the
Compton profile also scales, provided
that the potential energy due to the confinement dominates the Coulomb interaction.

\section{Excited states: Elitzur-Susskind bound state resonances}
\label{sec:susskind}
\begin{figure}[h]
\includegraphics[width = .49\textwidth]{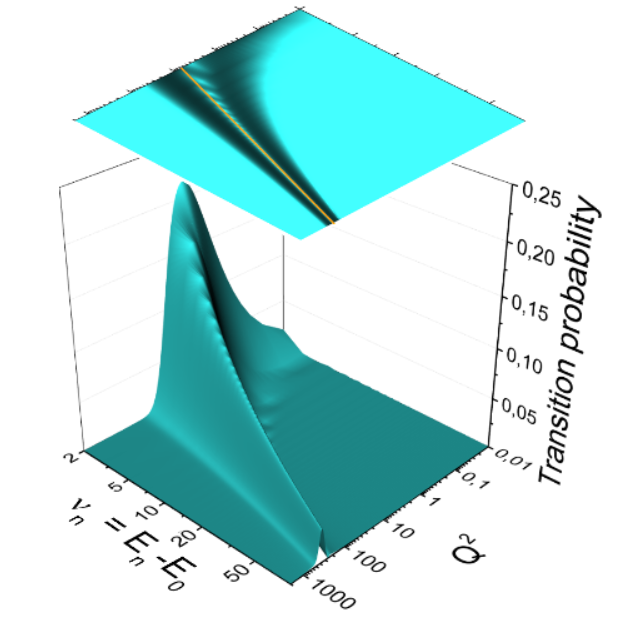}
\caption{Transition probability $T(\nu_n,Q^2)$ computed
for the confining potential with $\eta=2$ in
the presence of the Coulomb interaction. The scaling direction is seen by projection into 
the $(\nu_n,Q^2)$ plane. 
\label{graph2}}
\end{figure}

Elitzur and Susskind employed a simple confining potential~\cite{el.su.72}
to explain the scaling of the
resonance excitations in deep inelastic reactions~\cite{bl.gi.70,bjor.69}.
In their simplified model the probability for transitions between the bound states in the confining potential
was computed using the dipol approximation and was 
shown to be compatible with the scaling of resonant
excitations~\cite{el.su.72,elit.71,kogu.73} in deep inelastic reactions.
Contrary to Ref.~\cite{el.su.72} in which the WKB 
approximation was used, 
we solve eq.~\ref{eq:model1} numerically for a pair of point particles with masses
$m_1$, $m_2$ in a potential which is sufficiently deep
to bind states.
We assume that the momentum $Q$ is absorbed by one
of the particles, and the bound pair is lifted to
the $n^{th}$-level, at an energy $\nu_n = E_n - E_0$. 
Using the solutions of eq.~\ref{eq:model1}
we evaluate the  matrix elements $F\left(\nu_n,Q^2\right)$ for the transition into the $n^{th}$
bound level due to the absorption of a momentum ${\bf Q}$.
We note that the solutions of
eq.~\ref{eq:model1} produce bound states~\cite{Maha-2006} irrespective of the sign
of the confining potential. 
The transition probability is the square of
the transition matrix element:
 \begin{equation}
T\left(\nu_n,Q^2\right) 
= | \braket{\psi_{n}|e^{i{\bf Q}\cdot{\bf x}_1}|\psi_0} |^2   
=| \braket{\psi_{n}|e^{i{\bf Q}_1{\bf x}}|\psi_0} |^2 \ .
\end{equation}
Here $\psi_0$ and $\psi_{n}$ describe the ground state 
and the $n^{th}$ bound state of the confining potential,
respectively, with  
${\bf Q}_i={\bf Q}\, m_i/(\sum_i m_i)$. For electrons 
$m_1=m_2$
whereby ${\bf Q}_i={\bf Q}/2$ as the scaled momentum.
For finite $Q^2$ the transition probability
$T(\nu_n,Q^2)$ leads to $n$  discrete resonances.
The numerical results are presented in Fig.~\ref{graph2}. 

In Ref.~\cite{el.su.72} it was noted that the matching between the phase of $\psi_{n}$
and the exponential factor $e^{i{\bf Q}_1{\bf x}}$ implies a linear relation (``scaling direction'') between the square of the transferred momentum, $Q^2$, and the excitation energy $\nu_n$.
The scaling direction 
therefore represents a line in the $(\nu_n,Q^2)$-plane which we illustrate in the upper part of Fig.~\ref{graph2}.
For any other direction in the $(\nu_n,Q^2)$-plane the transition probability
decays exponentially (no phase matching).
This result was already derived in Ref.~\cite{el.su.72} within the WKB approximation, where the line has slope one.
It is interesting to note that this result
is not exactly reproduced in the present calculations, where the Coulomb interaction is taken into account.

\begin{figure}[htbp]
\includegraphics[width = .45\textwidth]{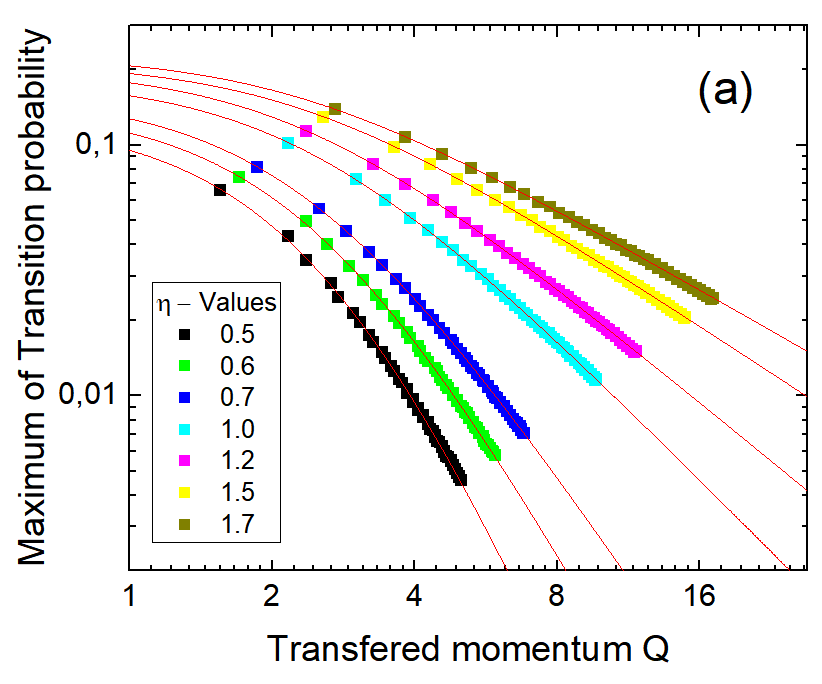}
\includegraphics[width = .45\textwidth]{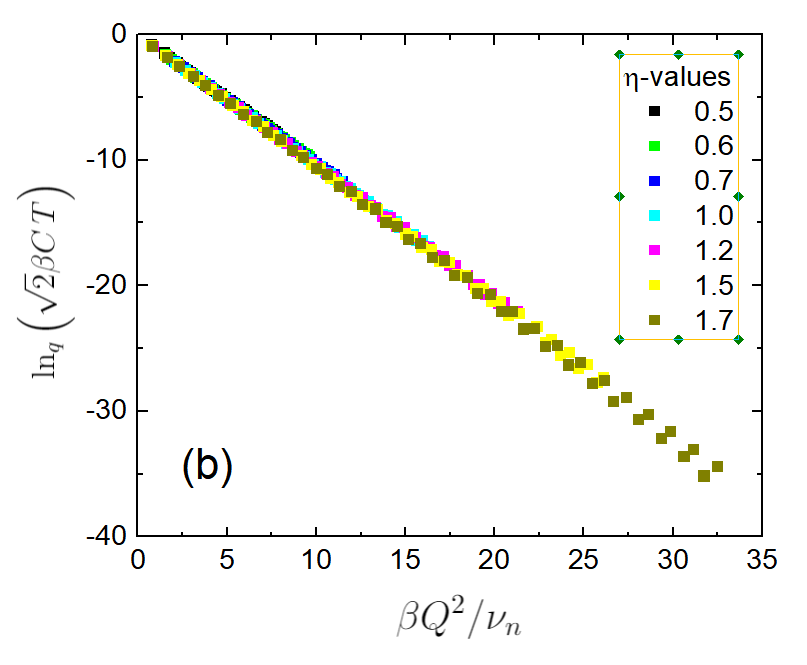}
\caption{(a)~The maxima of the transition probabilities corresponding to each transferred momentum Q are plotted for different confining potential strengths $\eta$. Dots are numerical data, while the red lines show the fit with a $q$-Gaussian distribution. (b) By scaling all maxima the transition probabilities collapse onto a single curve.
\label{graph4} }
\end{figure}

From the fact that the scaling direction is essentially a line
in the $(\nu_n,Q^2)$-plane along which the transition
probability is maximal, we can identify the ratio 
$Q^2/{\nu_n} \equiv \gamma$ as a scaling variable. 
We fitted the maxima along the 
scaling direction using a $q$-Gaussian
form:
\begin{equation}
 T(\nu_n,Q^2):= T_{q,\beta}(Q^2/\nu_n)= \frac{1}{\sqrt{2}\beta C_{q}} \exp_{q}\left(-\beta Q^2/\nu_n \right).  \label{Ttsal}
\end{equation}
In the limit of large momenta we find:
\begin{equation}
\lim_{Q^2\rightarrow \infty} T_{q,\beta}(Q^2/\nu_n) 
 \bigg|_{\nu_n=Q^2/\gamma} \propto \gamma^{2/(1-q)} \ ,
 \end{equation}
i.e. the $q$-Gaussian takes the form
of a power-law, which is characterized
by scale invariance. 
In Fig.~\ref{graph4}(a) we show the $q$-Gaussian fits to the
maxima of the transition probability for different confining
potentials, i.e., different values of $\eta$. In the limit of 
large momentum transfer, $Q^2\rightarrow \infty$, we obtain,
for all $\eta$ values, power laws along the 
``scaling direction" with a particular scaling exponent. 
Due to the scaling property they are all equivalent up to 
constant factors. This behavior is presented in 
Fig.~\ref{graph4}(b) where the $q$-logarithm of the transition
probability is plotted against the scaling variable $Q^2/\nu_n$. Here $\ln_q$ is the $q$-analog of the logarithm
defined by: $\ln_q x := (x^{1-q}-1)/(1-q)$. This plot is seen
to produce an essentially linear relation between the scaled
transition probabilities, which collapse onto a single curve.
Deviations are due to finite-size effects and numerical 
precision in the high $Q$-regime.

\begin{figure}[htbp]
\includegraphics[width = .45\textwidth]{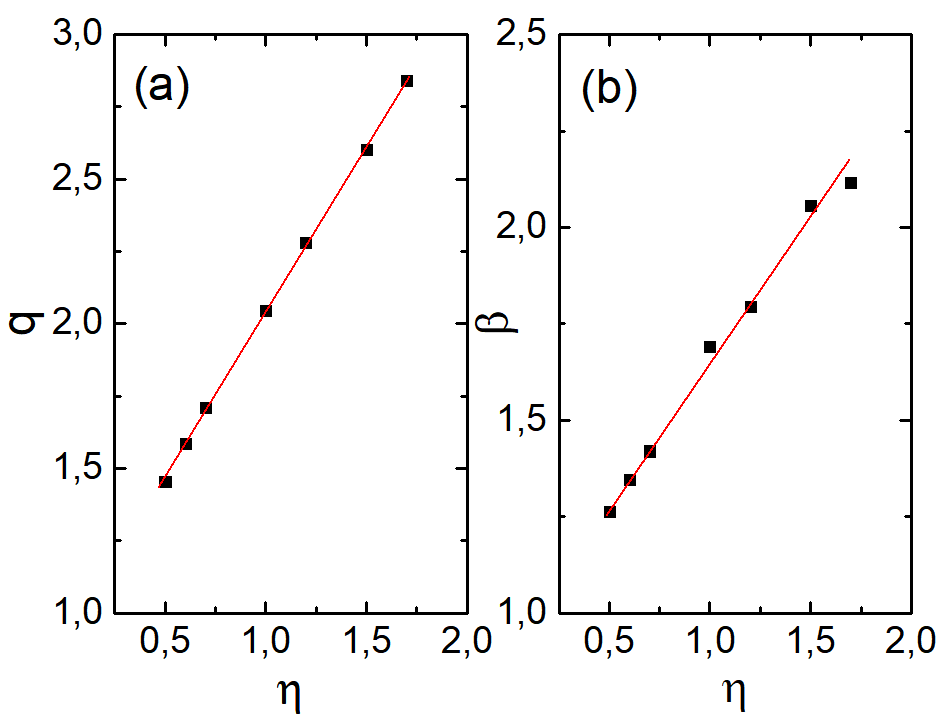}
\caption{(a)/(b) The dependence of $q$/$\beta$ for different $\eta$ values. The red line is a guide to the eye.
\label{graph5}}
\end{figure}

The $q$ and $\beta$ values for the fits in Fig.~\ref{graph4} 
are shown in fig.~\ref{graph5}(a),(b). We see that
$q$ and $\beta$ increase linearly for growing $\eta$ values.
With increasing $\eta$ the confining potential becomes
steeper as the interparticle distance increases. Therefore
one expects that the wave function (eigenfunction of $H$)
is more localized, which leads to a slower decay of the 
transition probabilities at high momenta.

\section{Discussion}
\label{sec:Discussion}
The results presented in this paper were initiated by the 
question whether, and under what conditions, the momentum 
distribution of a Coulomb systems shows scaling behavior. 
Following Kimball's approach~\cite{kimb.73,kimb.75}, we 
computed the momentum distribution of two interacting electrons
by numerically solving the two-particle Schr\"odinger equation
for a repulsive Coulomb interaction in the presence of a 
confining potential.
We found that $n({\bf k})$ can be parametrized by a
$q$-Gaussian in the entire momentum range. This crossover 
region connects the low-momenta region, described by an 
ordinary Gaussian momentum dependence, with a power-law
behavior at large momenta.
In the confinement dominated (intermediate) momentum
region we used the method of Elitzur and Susskind~\cite{el.su.72} and demonstrated that
bound-state resonances also show scaling behavior. 
In particular, we demonstrated that $q$-Gaussians are suitable
scaling functions for the maxima of the transition probability.
Indeed, the $q$-Gaussian behavior is expected to enter in this
investigation since it is the natural mathematical
function that can describe fat-tail distributions, whose 
asymptotic momentum dependence is not exponential but is
described, for example, by a power law.
Whenever the Coulomb interaction dominates the
confining potential (in the large momenta region) our results 
recover the exact analytic results obtained by renormalization
group techniques~\cite{an.bo.10,bo.ro.12,ho.ba.13}.
It would be desirable to gain insight into the numerically 
derived scaling properties also within an analytic approach.

Using density functional theory (DFT)~\cite{ho.ko.64,kohn.99,jo.gu.89,jone.15}
in combination with the impulse approximation
we recently showed that Compton profiles of the first column 
elements of the periodic table can be collapsed onto a single
curve~\cite{se.ap.18} which can be fitted by a $q$-Gaussian 
with element specific $(q,\beta)$-parameters. In that study we 
did not address the questions of why there should be scaling 
behavior at all, and why the $q$-Gaussian was found to be a
suitable scaling function.
In view of the fact that in the electronic band theory of solids the periodic ionic potential provides a natural confining potential, the results of the present paper may provide an explanation of the unexpected scaling behavior of the Compton profiles of the alkali elements~\cite{se.ap.18}.

For the application of the DFT~\cite{ho.ko.64,kohn.99,jo.gu.89,jone.15}
in the framework of the Kohn-Sham ansatz the
knowledge of the exchange-correlation functional is crucial.
A central quantity is the so-called coupling constant
integrated pair-correlation function. It accounts for
the electronic correlations contribution into the
kinetic energy~\cite{jo.gu.89} and is the input
to the derivation of the gradient corrected
functionals~\cite{pe.bu.96a,pe.bu.96b}.
The contribution of electronic correlations to the kinetic
energy has been analyzed also in momentum space~\cite{go.pe.02}, and the
limits of large and small momenta were discussed and found to
be in agreement with results of Kimball~\cite{kimb.73,kimb.75}.
In fact, according to our investigation which extends the work
of Kimball by including confining potentials, a $q$-Gaussian
fit for the momentum density is possible for every value of the
momentum, and the somewhat arbitrary separation between 
short- and long-range decomposition can be avoided.
We expect that the concept of scaling of the momentum distribution in terms of a $q$-Gaussian can be further developed such that it actually provides new exchange-correlation functionals.

\section{acknowledgments}
Financial support by the Deutsche Forschungsgemeinschaft through TRR80 (project F6)
Project number 107745057 and useful discussions with M. Sekania and M. Kollar are gratefully acknowledged.

\bibliography{paper}

\end{document}